\documentclass[sigconf]{acmart}
\usepackage{algorithmic}
\usepackage{graphicx}
\usepackage{textcomp}

\usepackage{enumitem}

\usepackage{todonotes}
\usepackage{url}
\usepackage{booktabs}
\usepackage{caption}
\usepackage{multirow}
\usepackage{censor}
\usepackage{hyperref}
\usepackage{soul}

\usepackage{subcaption}
\usepackage{mathrsfs}
\usepackage{numprint}
\npdecimalsign{,}
\nprounddigits{2}
\usepackage{colortbl}

\usepackage[draft]{minted}

\definecolor{codegreen}{rgb}{0.0, 0.5, 0.0}

\setminted[python]{ %
    linenos=false,             % Line numbers
    autogobble=true,          % Automatically remove common white space
    frame=lines,
    framesep=2mm,
    fontsize=\footnotesize
    }
    
\definecolor{main}{HTML}{5989cf}    % setting main color to be used
\definecolor{sub}{HTML}{cde4ff}     % setting sub color to be used

\usepackage[many]{tcolorbox}    	% for COLORED BOXES (tikz and xcolor included)

\newtcolorbox{RQanswer}[1][]{%
    colback=sub,
    colframe=black!5,
    notitle,
    sharp corners,
    borderline west={2pt}{0pt}{main!80!black},
    enhanced,
    breakable,
    right=6pt,
    top=0pt,
    bottom=0pt,
    }
    
\usepackage{breqn}
\usepackage{amsthm}
\newtheorem{definition}{Definition}

\usepackage{tablefootnote}

\copyrightyear{2024} 
\acmYear{2024} 
\setcopyright{rightsretained} 
\acmConference[ICSE '24]{2024 IEEE/ACM 46th International Conference on Software Engineering}{April 14--20, 2024}{Lisbon, Portugal}
\acmBooktitle{2024 IEEE/ACM 46th International Conference on Software Engineering (ICSE '24), April 14--20, 2024, Lisbon, Portugal}\acmDOI{10.1145/3597503.3639133}
\acmISBN{979-8-4007-0217-4/24/04}

\begin{document}

\title{Traces of Memorisation in Large Language Models for Code\\
% \thanks{Applicable funding}
}

% \author{
% \IEEEauthorblockN{Ali Al-Kaswan}
% \IEEEauthorblockA{\textit{Delft University of Technology} \\
% % \textit{name of organization (of Aff.)}\\
% Delft, The Netherlands \\
% \email{a.al-kaswan@tudelft.nl}} 
% \and
% \IEEEauthorblockN{Maliheh Izadi}
% \IEEEauthorblockA{\textit{Delft University of Technology} \\
% % \textit{name of organization (of Aff.)}\\
% Delft, The Netherlands \\
% \email{m.izadi@tudelft.nl}}
% \and
% \IEEEauthorblockN{Arie van Deursen}
% \IEEEauthorblockA{\textit{Delft University of Technology} \\
% % \textit{name of organization (of Aff.)}\\
% Delft, The Netherlands \\
% \email{arie.vandeursen@tudelft.nl}}
% }
\author{Ali Al-Kaswan}
\email{a.al-kaswan@tudelft.nl}
\affiliation{%
  \institution{Delft University of Technology}
  \city{Delft}
  \country{The Netherlands}
}
\author{Maliheh Izadi}
\email{m.izadi@tudelft.nl}
\affiliation{%
  \institution{Delft University of Technology}
  \city{Delft}
  \country{The Netherlands}
}
\author{Arie van Deursen}
\email{arie.vandeursen@tudelft.nl}
\affiliation{%
  \institution{Delft University of Technology}
  \city{Delft}
  \country{The Netherlands}
}

\begin{abstract}
Large language models
have gained significant popularity because of their ability to generate
human-like text and potential applications in various fields,
such as Software Engineering. 
Large language models for code are commonly
trained on large unsanitised corpora of source code scraped from
the internet. 
The content of these datasets is memorised and
can be extracted by attackers with 
data extraction attacks.
%our contribution
In this work, 
we explore memorisation 
in large language models for code
and compare the rate of memorisation 
with large language models trained on natural language.
We adopt an existing benchmark for natural language 
and construct a benchmark for code 
by identifying samples that are vulnerable to attack.
We run both benchmarks against a variety of models,
and perform a data extraction attack.
%findings
We find that 
large language models for code
are vulnerable to data extraction attacks,
like their natural language counterparts.
From the training data 
that was identified to be potentially extractable
we were able to extract 47\% from 
a CodeGen-Mono-16B code completion model.
We also observe that models memorise more,
as their parameter count grows,
and that their pre-training data
are also vulnerable to attack.
We also find that
data carriers are memorised at a higher rate than regular code or documentation
and that different model architectures memorise different samples.
%future work
Data leakage has severe outcomes, 
so we urge the research community to 
further investigate the extent of this phenomenon using a wider range of models 
and extraction techniques 
in order to build safeguards 
to mitigate this issue.
\end{abstract}

\begin{CCSXML}
<ccs2012>
   <concept>
       <concept_id>10002978</concept_id>
       <concept_desc>Security and privacy</concept_desc>
       <concept_significance>300</concept_significance>
       </concept>
   <concept>
       <concept_id>10011007</concept_id>
       <concept_desc>Software and its engineering</concept_desc>
       <concept_significance>500</concept_significance>
       </concept>
   <concept>
       <concept_id>10010147.10010257</concept_id>
       <concept_desc>Computing methodologies~Machine learning</concept_desc>
       <concept_significance>500</concept_significance>
       </concept>
 </ccs2012>
\end{CCSXML}

\ccsdesc[300]{Security and privacy}
\ccsdesc[500]{Software and its engineering}
\ccsdesc[500]{Computing methodologies~Machine learning}
\keywords{Large Language Models, Privacy, Memorisation, Data Leakage}

\maketitle

\section{Introduction}
\label{introduction}
In recent years, Large Language Models (LLMs) have garnered considerable interest in the realm of Natural Language Processing (NLP) owing to their exceptional accuracy in performing a broad spectrum of NLP tasks~\cite{lu2021codexglue}. These models, trained on extensive amounts of data, exhibit increased accuracy and emergent abilities as their parameter count grows from millions to billions~\cite{xu2022systematic}. LLMs designed for coding are also trained on vast amounts of data and can effectively learn the structure and syntax of programming languages. As a result, they are highly adept at tasks like generating~\cite{fried2022incoder}, summarising~\cite{alkaswan2023extending}, and completing code~\cite{izadi2022codefill}.

Large language models also exhibit emergent capabilities~\cite{wei2022emergent}. These abilities cannot be predicted by extrapolating scaling laws and only emerge at a certain critical model size threshold~\cite{wei2022emergent}. This makes it appealing to train ever-larger models, as capabilities such as chain-of-thought prompting~\cite{wei2022chain} and instruction tuning~\cite{ouyang2022training} only become feasible in models with more than 100B parameters~\cite{wei2022emergent}.

Many have noted that large language models trained on natural language are capable of memorising extensive amounts of training data~\cite{alkaswan2023abuse, carlini2022membership, carlini2022quantifying, carlini2022privacy, henderson2023foundation, sun2022coprotector, tirumala2022memorization, karmakar2022codex, biderman2023emergent, feldman2020does, chen2023overconfidence, lukas2023analyzing, ishihara2023training}.

The issue of memorisation in source code is distinct from that of natural language. Source code is governed by different licences that reflect different values than natural language~\cite{henderson2023foundation, Choksi2023WhoseTI}. 
Hence, in addition to privacy considerations, the memorisation of source code can have legal ramifications. 
The open-source code used in LLM training for code is frequently licenced under nonpermissive copy-left licences, such as GPL or the CC-BY-SA licence employed by StackOverflow~\cite{alkaswan2023abuse}.\footnote{StackOverflow Licence: https://stackoverflow.com/help/licensing} 
Reusing code covered by these licences without making the source code available under the same licence is considered a violation of copyright law. In some jurisdictions, this leaves users of tools such as CoPilot at legal risk~\cite{alkaswan2023abuse, henderson2023foundation, Choksi2023WhoseTI}. 
Licences are unavoidably linked to the source code, as they enforce the developers' commitment to sharing, transparency, and openness~\cite{Choksi2023WhoseTI, alkaswan2023abuse}. Sharing code without proper licences is also ethically questionable~\cite{alkaswan2023abuse, sun2022coprotector, henderson2023foundation}. 

Memorised data can also include private information~\cite{carlini2021extracting, carlini2023extractingDiffusion, ippolito2022preventing}. These privacy concerns extend to code, which can contain credentials, API keys, emails, and other sensitive information as well~\cite{alkaswan2023abuse, basak2023secretbench}. Memorisation could therefore put the private information contained in the training data at risk. 

Recently, attacks which leverage memorisation have successfully extracted (or reconstructed) training data from LLMs~\cite{carlini2021extracting, alkaswan2023Targeted, biderman2023emergent, ishihara2023training}. The US National Institute of Standards and Technology (NIST) considers data reconstruction attacks to be the most concerning type of privacy attack against machine learning models~\cite{oprea2023adversarial}. OWASP classifies Sensitive Information Disclosure (LLM06) as the sixth most critical vulnerability in LLM applications.~\footnote{OWASP Top 10 for Large Language Model Applications: \url{https://owasp.org/www-project-top-10-for-large-language-model-applications/}}

Larger models are more likely to memorise more data and are more vulnerable to data extraction~\cite{carlini2021extracting, biderman2023emergent, oprea2023adversarial, ishihara2023training}. The effort to create ever larger LLMs, therefore, creates models which carry more risk.

To our knowledge, previous studies have investigated data memorisation and extraction attacks in natural language, but there has been no empirical investigation of LLMs for code. In this work, we investigate to which extent large language models for code memorise their training data and how this compares to memorisation in large language models trained on natural language. There is no comprehensive framework or approach for measuring memorisation.

% Experiments
We start by defining a data extraction security game that is grounded in the theory behind membership inference attacks and the notion of k-extractability. Using this security game we define a framework to quantify memorisation in LLMs. We use data extraction as an estimator of memorisation. While memorisation of training data can manifest in the form of non-exact duplication, measuring the rate of data extraction data extraction provides a lower bound of memorisation in a model.

We perform experiments leveraging the SATML training data extraction challenge, an existing dataset for natural language.\footnote{Language Models Training Data Extraction Challenge: \url{https://github.com/google-research/lm-extraction-benchmark}} We extend this benchmark by testing memorisation on more models. 

We construct a similar dataset for code, by mining data from the Google BigQuery GitHub dataset and by using a CodeGen code generation model~\cite{Nijkamp2022CG}. Similarly to the natural language dataset, we first identify samples vulnerable to attack to build a benchmark. We then tested a variety of models on this benchmark. We finally compare the rate of memorisation between text and code models.

% Key result:
Our key result: \textit{Large language models trained on code memorise their training data like their natural language counterparts and are vulnerable to attack.}
%
% We find that source code pre-trained language models memorise training data at a lower rate than natural language pre-trained models. We also observe that the models memorise more and become more vulnerable as their parameter count grows. Finally, we observe that code models are vulnerable to attacks on their pre-training data.
%
% Main contributions:
To summarise, the main contributions of this paper are:
\begin{itemize}
    % \item An empirical evaluation demonstrating memorisation in LLMs for natural language
    % \item An empirical evaluation demonstrating memorisation in LLMs for source code
    % \item A novel benchmark to measure memorisation in code models 
    \item A novel approach, using a data extraction security game, to quantify memorisation rates of code or natural language models
    \item A benchmark of key memorisation characteristics for 10 different models of different sizes
    \item An empirical assessment of memorisation in code models demonstrating that (1) code models memorise training data, albeit at a lower rate than natural language models; (2) larger models, with more parameters, exhibit more memorisation; (3) data carriers (such as dictionaries) are memorised at a higher rate than, e.g., regular code, documentation, or tests; (4) different model architectures memorise different samples. 
    \item We make the code to run the evaluation available to allow others to replicate our results and to evaluate other models.~\footnote{GitHub repo: \url{https://github.com/AISE-TUDelft/LLM4Code-extraction}}
\end{itemize}

\section{Background and Related Work}
\label{memorisation}

% \begin{todolist}
%     \item[\done] Define memorisation
%     \item[\done] explain MIA 
%     \item[\done] explain guided and unguided data extraction attacks
% \end{todolist}

\subsection{Memorisation}
In the context of language models, memorisation refers to the ability of a model to remember and recall specific details of the data it has been trained on. This occurs when a model overfits the training data, meaning it becomes overly specialized and fails to generalise well to new or unseen data~\cite{feldman2020does, choquette2021label}. As a result, the model can accurately recall specific phrases, sentences, or even entire documents from the training data. Besides the privacy concerns explained in~\autoref{introduction}, memorisation also causes an overestimation of performance. It has, for instance, been observed that CodeX can complete HackerRank problems without receiving the full task description~\cite{karmakar2022codex}.

While memorisation can lead to high accuracy, it is not necessarily an indication of good generalisation performance. A model that has memorised the training data may struggle to perform well on new or unseen data, leading to poor performance in real-world applications. Additionally, memorisation can reduce the ability of the model to adapt its output to specific use cases. For example, when slightly changing HackerRank problems, CodeX~\cite{chen2021evaluating} struggles to produce a correct solution, instead regurgitating solutions for the original problem~\cite{karmakar2022codex, tan2023dimensionality}.

%Memorisation can in some cases be beneficial. For data with a long-tailed distribution, such as large mined corpora of text and code, the model needs to memorise outliers to obtain optimal performance~\cite{feldman2020does}. 

\subsection{Membership Inference Attacks}
Membership inference attacks are a type of attack that aims to determine whether a specific data point was included in the training data of a machine learning model. The goal of these attacks is to infer whether a given data point was used to train the model or not, without having access to the training data itself.

The first membership inference attack against machine learning models was proposed by~\citeauthor{shokri2017membership} to target classification models deployed by Machine Learning as a Service (MLaaS) providers~\cite{shokri2017membership}. Since then the field has expanded and attacks have been proposed that target generative models~\cite{hilprecht2019monte} and LLMs~\cite{hisamoto2020membership}. Recently, membership inference attacks have been proposed against transformer-based image diffusion models such as Stable Diffusion~\cite{dubinski2023towards}.

We refer to the security game defined by~\citeauthor{carlini2022membership}~\cite{carlini2022membership} to define a membership inference attack in~\autoref{def:mia}. In this game, the adversary wins if they have a non-negligible advantage \( > \frac{1}{2} + \epsilon\). In simpler terms, the adversary needs to be able to distinguish between data that was included and which was not included in the training data for a given model, while only being allowed query access to the model and data distribution. 

Membership inference attacks are primitive for measuring the leakage of a machine learning model and are often a starting point for more extensive attacks~\cite{hu2022membership, carlini2022membership, mireshghallah2022quantifying}. While membership inference is a weaker privacy violation than memorisation, the National Institute of Standards and Technology (NIST) still considers membership inference to be a violation of the confidentiality of training data~\cite{hu2022membership}.

\begin{definition}[Membership inference security game~\cite{carlini2022membership}]
\label{def:mia}
The game proceeds between a challenger $\mathcal{C}$, an adversary $\mathcal{A}$, a data distribution $\mathbb{D}$ and a model $f$:
\begin{enumerate}
    \item The challenger samples a training dataset $D \gets \mathbb{D}$ and trains a model $f_\theta \gets \mathcal{T}(D)$ on the dataset $D$.
    \item The challenger flips a bit $b$, and if $b=0$, samples a fresh challenge point from the distribution $(x, y) \gets \mathbb{D}$ {(such that $(x, y) \notin D$)}. Otherwise, the challenger selects a point from the training set $(x, y) \gets D$.
    \item The challenger sends $(x, y)$ to the adversary.
    \item The adversary gets query access to the distribution $\mathbb{D}$, and to the model $f_\theta$, and outputs a bit $\hat{b}$
    \item Output 1 if $\hat{b}=b$, and 0 otherwise.
\end{enumerate}
\end{definition}

\subsection{Data Extraction Attacks}
Data extraction attacks are a stronger type of attack where an adversary extracts a data point used to train a model. Attacks can be divided into two types for LLMs, namely guided and unguided attacks~\cite {alkaswan2023Targeted}.

In an unguided attack, the adversary does not know the sample to be extracted from the model. The adversary simply attempts to extract any training point, contained anywhere in the training corpus~\cite{carlini2021extracting, carlini2023extractingDiffusion, oh2023membershipKo, carlini2022privacy}.

In this work, we focus on targeted attacks. In a targeted attack, the adversary is provided with a prefix, which is the first half of the sequence and is then tasked with recovering the suffix, which is the second half of the sequence. Targeted attacks are more security-critical as they allow the targeting of specific information, such as the extraction of emails~\cite{alkaswan2023Targeted, carlini2023extractingDiffusion,henderson2023foundation,huang2022large, mireshghallah2022quantifying}.

We ground our definition of memorisation and extractability in the definition of k-extractability provided by~\citeauthor{biderman2023emergent}, which was originally inspired by the framework of k-eidetic memorisation introduced by \citeauthor{carlini2021extracting}~\cite{carlini2021extracting}.
\begin{definition}[k-extractability~\cite{biderman2023emergent}]
\label{def:kex}
A string s is said to be k-extractable if it (1) exists in the training data, and (2) is generated by the language model by prompting with k prior tokens.
\end{definition}

\subsection{Natural Language Dataset}
The dataset used for the attack on natural language models is provided by the SATML'23 Language Model Data Extraction Challenge\footnote{Language Models Training Data Extraction Challenge: \url{https://github.com/google-research/lm-extraction-benchmark}\label{lmext}}. The dataset consists of 15K training, 1K validation, and 1K test samples. The test samples were not released and were only used by the competition organisers. Each sample is divided into a 50-token prefix and a 50-token suffix. For our evaluation, we use the validation set.\footref{lmext}

The participants had to use a GPT-NEO 1.3B model to extract the suffix using the prefix. The winning entry prompted the model with the prefix, extracted 100 suffixes for each prefix, and trained a binary classifier to select the most correct suffix~\cite{alkaswan2023Targeted}. 

The dataset was constructed by analysing Pile~\cite{pile}, which is the corpus used to train the GPT-NEO family of models~\cite{gpt-neo}. The Pile is an 825GB English language dataset, which itself consists of 22 high-quality sub-datasets, ranging from books, academic papers and even code~\cite{pile}. The Pile was constructed to improve the cross-domain applicability of LLMs. The Pile~\cite{pile} is also used as a pretraining dataset for a variety of code models~\cite{alkaswan2023abuse}.\footnote{Following a DMCA takedown request against the Books3 subset of the Pile, as of December 2023 the Pile is no longer publically available: \url{https://archive.ph/1h00A}}

The organisers extracted all the unique 150 token sequences from the 800GB corpus. Sequences were filtered to include only those that are duplicated at least 5 times. They were then split into a pre-prefix, prefix, and suffix, each 50 tokens long. The GPT-NEO model was then prompted with the pre-prefix and prefix (100 tokens). If the model produces the suffix, using greedy decoding, the sequence is considered extractable. The challenge dataset was constructed from the extractable sequences and only includes the prefix and suffix.\footref{lmext}

\section{Approach}
\label{approach}
To measure memorisation in LLMs4Code we first formally define a data extraction game and we construct a dataset of code samples.

\subsection{Data Extraction Security Game}
We consider the models as black-box systems. We define a security game inspired by the membership inference attack security game in~\autoref{def:mia} and the notion of k-extractability in~\autoref{def:kex}:

\begin{definition}[Data extraction security game]
\label{def:ext}
Given a challenger $\mathcal{C}$, an adversary $\mathcal{A}$, a data distribution $\mathbb{D}$ and a model $f$ the game is defined as follows:
\begin{enumerate}
    \item The challenger samples a training dataset $D \gets \mathbb{D}$ and trains a model $f_\theta \gets \mathcal{T}(D)$ on the dataset $D$.
    \item $\mathcal{C}$ samples a sample \(D_n = (p, s)\) where \(D_n \in D\). The prefix \(p\) is provided to the adversary $\mathcal{A}$.
    \item $\mathcal{A}$ is allowed query access to the model \(f_\theta\) and may perform any other polynomial-time operations
    \item $\mathcal{A}$ outputs his prediction sequence \(\hat{s}\)
    \item If \(\hat{s} = s\), $\mathcal{A}$ wins, otherwise $\mathcal{C}$ wins
\end{enumerate}
\end{definition}

In other words, given a prefix (1), the adversary is challenged to extract the correct suffix in the training data from the model. The adversary can query the model (2), but has no access to the weights, unlike the game proposed by~\citeauthor{alkaswan2023Targeted}~\cite{alkaswan2023Targeted}. The adversary then predicts the suffix (3) and wins if it matches the actual suffix in the training data. 
% \todo[inline]{Maybe explain that p , s in training data but p , s' where s' =/ s not in training data in the original attack. But we cannot give this guarantee}

There are some difficulty modifiers to adjust the difficulty of the challenge:
\begin{enumerate}
    \item The selection of the dataset \(D \subset \mathbb{D}\). As observed by previous works, not all training samples are as hard to extract as others. In particular, samples that are highly duplicated\footref{lmext} or outliers~\cite{carlini2022privacy} are more vulnerable to attack.
    \item The choice of model \(M_\theta\). Some models are more likely to memorise samples than others, namely larger models have been observed to memorise more samples~\cite{carlini2022quantifying, carlini2021extracting, carlini2023extractingDiffusion, brown2022does, biderman2023emergent, ishihara2023training}.
    \item The length of the prefix \(p\). It has been found that longer prefixes elicit more memorisation\footref{lmext}~\cite{carlini2022quantifying, carlini2021extracting, ishihara2023training}. Note that this length is equivalent to the \(k\) in definition \autoref{def:kex}.
    \item The victory condition \(\hat{s} = s\), instead of targeting verbatim memorisation, a fuzzy match could also be considered~\cite{ishihara2023training}. 
\end{enumerate}
In this work, we take inspiration from the competition organised by Carlini et al. and use modifiers (1) and (3) to construct a set of extractable samples. We shorten the prefix of the extractable samples and use this set of hard but extractable samples to perform an evaluation on different models (2). We also measure fuzzy match scores (4) and compare them with the extract match rate.

\subsection{Code Dataset Construction}
To measure the memorisation in LLMs for code, we first need to construct a dataset similar to the one used in the SATML'23 Language Model Data Extraction Challenge. As there is no code benchmark available, we build one from scratch. This presents several challenges:

Firstly, for some code models, the training data is not published by the authors, which makes it impossible to determine what data were included in the training of these models. We must therefore experimentally determine which data points were presumably included in the training data for each of the models. 
This has implications for the transferability of the benchmark set, as the training data might differ for each model. Not all models are trained in all programming languages as well, so we must select a common language to test multiple models.

Secondly, since all publicly available code is potentially part of the training data, the search space for extractable data points is massive. 

We limit our evaluation to Python since we found that the vast majority of models support Python and have some Python in their training corpus. We source the potentially memorised data from GitHub. We mine Python files using the Google BigQuery Github dataset.~\footnote{GitHub on BigQuery: \url{https://cloud.google.com/blog/topics/public-datasets/github-on-bigquery-analyse-all-the-open-source-code}}

We filter the files to include only nonbinary files longer than 150 tokens. We only consider files that have five or more duplicates on GitHub and randomly select 150 token spans from anywhere in the file. Similarly to the natural language dataset~\footnote{Language Models Training Data Extraction Challenge: \url{https://github.com/google-research/lm-extraction-benchmark}}, we split the 150 token span into a pre-prefix, a prefix, and a suffix, each 50 tokens long. We prompt a CodeGen-2B-Mono model~\cite{Nijkamp2022CG} with the pre-prefix and prefix. We select this model because it is decently sized (there are smaller and larger variants of the model), it is specifically trained on Python and it is the highest performing publically-available model for the Human-Eval benchmark~\cite{Nijkamp2022CG}.

If the model can predict the suffix, with the 100-token prompt, we consider the sample to be extractable. We randomly select 1K extractable samples to perform our evaluation. We construct the dataset from the prefixes and suffixes. 

Our dataset construction procedure differs from the procedure used by Carlini et al. in one aspect. Our dataset does not guarantee that for every \(D_n = (s,p)\) there does not exist a \((s,p') \in D\) where \(p \neq p'\). There are two main reasons for omitting this step:
\begin{itemize}%[leftmargin=*]
    \item For many models in our evaluation we do not have access to the training data and possible pre-training data. The organisers could guarantee that the model under investigation was only exposed to the Pile. We want our approach to work for settings in which the investigator has no access to the training data.
    \item The computational cost of identifying all unique samples \(D_n = (s,p)\) is extremely large for a dataset of this size and our aim is to create an approach that does not require such enormous compute capabilities.
\end{itemize}

\section{Methodology}
\label{methodology}
\subsection{Research Questions}
\begin{itemize}
    \item[RQ1:] \textit{How does the rate of memorisation compare between Natural Language and Code trained LLMs?} To compare the rate of memorisation, we run both the attack on natural language as well as code models and compare the results. Intuitively we expect code models to be able to memorise more since code is more structured and there is much more natural language data available. 
    \item[RQ2:] \textit{What type of data are memorised by code-trained LLMs?} We want to know if there is a code pattern that is memorised. To do this we take the set of samples vulnerable to attack and we manually analyse them by constructing a classification of the samples. 
    \item[RQ3:] \textit{How much overlap is there between the memorised samples in different code-trained LLMs?} Do some models memorise different samples than others? Could we perhaps leverage a selection of different models to extract more data and do some models memorise more of a certain type of sample than others? 
    \item[RQ4:] \textit{To what extent do LLMs trained in code leak their pre-training data?} Finally, we want to see if pre-trained models can also leak their pre-training data. To investigate this, we select a code model that has been pre-trained on the Pile and perform the natural language attack. We compare the performance of the original base model with that of the code-trained model to see how much training data is retained. When referring to a base model in this paper, we only mean models that were initialised with the \textbf{architecture} and \textbf{weights} of a different model. 
\end{itemize}

\subsection{Models}
The models, their developers, and their respective sizes are shown in~\autoref{tab:models}. We limit our evaluation to left-to-right autoregressive models, which are available on the HuggingFace Hub. 

\begin{table}
    \centering
    \caption{Natural language (top 4 rows) and code models under investigation}
    \begin{tabular}{ll|r}
    \noalign{\smallskip}\toprule
    Model           & Developers    &   Parameters              \\ 
    \cmidrule{1-3}
    GPT-NEO         & EleutherAI    &   125M, 1.3B, 2.7B        \\
    GPT-2           & OpenAI        &   117M, 345M, 774M, 1.5B  \\
    Pythia          & EleutherAI    &   70M, 160M, 410M, 1B     \\
                    &               &   1.4B, 2.8B, 6.9B        \\ 
    CodeGen-NL      & Salesforce    &   350M, 1B, 3B, 7B, 16B   \\
    \cmidrule{1-1}
    CodeGen-Mono    & Salesforce    &   350M, 1B, 3B, 7B, 16B   \\
    CodeGen-Multi   & Salesforce    &   350M, 1B, 3B, 7B, 16B   \\
    CodeGen2        & Salesforce    &   1B, 3.7B, 16B           \\
    CodeParrot      & Huggingface   &   110M, 1.5B              \\
    InCoder         & Facebook      &   1.5B                    \\
    PyCodeGPT       & Microsoft     &   110M                    \\
    GPT-Code-Clippy & CodedotAI     &   125M                    \\
    \noalign{\smallskip}\bottomrule
    \end{tabular}
    \label{tab:models}
\end{table}

For natural language evaluations, we used GPT-NEO~\cite{gpt-neo}, the models used to build the natural language dataset~\footref{lmext}. We select GPT-2~\cite{radfordlanguage_gpt2} to test the transferability of the prompts to a model trained on a different corpus. GPT-2 is trained on the WebText corpus, which was mined by finding all the outlinks on Reddit with more than 3 karma. We also investigate the Pythia~\cite{biderman2023pythia} suite of models, which are trained on the Pile~\cite{pile}. 

The CodeGen suite of models~\cite{Nijkamp2022CG} features a number of different models in a variety of sizes. The models were initialised and first pre-trained on the Pile; these models are the CodeGen-NL models. The CodeGen-NL models are then further trained on a dataset containing multiple programming languages to create the CodeGen-Multi models. The Multi models were finally trained on a dataset consisting of only Python code to create the CodeGen-Mono models. The CodeGen2 and Incoder models are both designed for infilling but have autoregressive capabilities as well~\cite{Nijkamp2022CG, fried2022incoder}. CodeParrot is a pre-trained GPT-2 model fine-tuned on the APPS dataset~\cite{sharma2023stochastic}. PyCodeGPT is a small and efficient code generation model based on the GPT-NEO architecture~\cite{CERT}. GPT-Code-Clippy is a pre-trained GPT-NEO model fine-tuned on code.

\subsection{Categorisation}
We build a classification of the 1K extractable 150-token samples by doing an explorative study. We find the following categories and classify each of the samples into one category. For simplicity, we classify each sample which has two purposes, into its majority category. The different categories are shown in \autoref{tab:cats}. We identified 5 different categories as shown in~\autoref{tab:cats}.

\begin{table}
    \centering
    \caption{Categories of memorised samples}
    \begin{tabular}{ll|r}%r}
    \noalign{\smallskip}\toprule
    Category        & Purpose               & Count     \\% &  Example          \\ 
    \cmidrule{1-3}
    Code            & Code Logic            & 679     \\%   & \tablefootnote{\href{https://github.com/securesystemslab/zippy/blob/ff0e84ac99442c2c55fe1d285332cfd4e185e089/zippy/lib-python/3/profile.py\#L406-L422}{profile.py}} \\
    Testing         & Test Code             & 87        \\%   & \tablefootnote{\href{https://github.com/pyparallel/pyparallel/blob/11e8c6072d48c8f13641925d17b147bf36ee0ba3/Lib/ctypes/test/test_win32.py\#L70C11-L81}{test\_win32.py}} \\
    License         & Licence information   & 13         \\%  & \tablefootnote{\href{https://github.com/inveniosoftware-attic/invenio-mobile/blob/1fc225debf62ef32ee6a7b70669ad3b3555ba9d4/index.html\#L1C5-L12}{index.html}} \\
    Docs            & Documentation         &  86        \\%  & \tablefootnote{\href{https://github.com/twisted/twisted/blob/c462c5f7d940260ea3ca553c8653d0d227bbf468/src/twisted/web/http.py\#L1886-L1895}{http.py}} \\
    Dicts           & Dictionaries or other data carriers & 135        \\%  & \tablefootnote{\href{https://github.com/mhammond/pywin32/blob/ba25e7533a6b20d4e263159e1f8d6b15483ba36d/win32/Lib/win32con.py\#L4660-L4674}{win32con.py}}  \\
    \noalign{\smallskip}\bottomrule
    \end{tabular}
    \label{tab:cats}
\end{table}

\subsection{Extraction}
We prompt the model under investigation with the prefix. We use the standard generation pipeline and the default generation configuration of the model as defined in the model configuration. For models which use a different tokeniser than the CodeGen tokeniser used for the dataset construction. We simply tokenise the sample again using the new tokeniser. Any samples that are too short under the new tokeniser are discarded. 

\subsection{Evaluation Metrics}
The models are prompted in a one-shot fashion with greedy decoding. We measure the exact match rate (EM). Additionally, we also measure the fuzzy match, using the BLEU-4 score. 
% Generally, a high BLEU-4 and relatively low EM score indicate good generalisation but low memorisation. Higher EM and lower BLEU-4 indicate more memorisation and less generalisation. We report the average Exact Match rates and average BLEU4 scores. 
For the model size, we measure the total parameter count. 

\subsection{Configuration}
We process and visualise the data with Modin $0.20.0$ and Pandas $2.0.1$. We run inference using Transformers version $4.16.2$ running on Torch $1.9.0$+cu$111$. The experiments were conducted on a cluster running RedHat $7$, we allocated $8$ CPU cores with $32$GB of RAM and an Nvidia A40 GPU with $48$GB of video memory. The GPU is running Nvidia driver version $530.30.02$ with Cuda $12.1$. 

For replication purposes, we only consider models that are runnable on our hardware. We found that the limitation was the GPU memory, so there are some models that we did consider but did not fit the GPU memory (such as InCoder-6.7B and StarCoder-base).

\section{Results}
\label{results}
We present the results of our experiments to answer the research questions, results are grouped per research question. 

\subsection{Natural Language vs Code}
The results of the attack are shown in~\autoref{tab:NL_res}. We found that we are able to extract 56\% of the samples with the largest GPT-NEO model. The medium-sized model, which was used to construct the dataset, achieved an exact match rate of 46\%. The models which were not trained on the Pile~\cite{pile} did not memorise much if any of the samples.

\begin{table}
    \centering
    \caption{Code attack performance on Large Language Models for Code}
    \begin{tabular}{ll|ll}
    \noalign{\smallskip}\toprule
    &                   &           \multicolumn{2}{c}{Memorisation rate} \\
    \cmidrule{3-4}
    Model                   & Parameters (M)    & EM        & BLEU-4    \\ 
    \cmidrule{1-4}
    CodeGen-350M-Mono       & 357               &   0.101   &   0.567   \\ 
    CodeGen-2B-Mono         & 2779              &   0.303   &   0.712   \\ 
    CodeGen-6B-Mono         & 7074              &   0.382   &   0.756   \\ 
    CodeGen-16B-Mono        & 16032             &   0.471   &   0.801   \\
    \cmidrule{1-1}
    CodeGen-350M-Multi      & 357               &   0.100   &   0.536   \\ 
    CodeGen-2B-Multi        & 2779              &   0.204   &   0.628   \\ 
    CodeGen-6B-Multi        & 7074              &   0.258   &   0.659   \\ 
    CodeGen-16B-Multi       & 16032             &   0.297   &   0.695   \\
    \cmidrule{1-1}
    CodeGen-2B-nl           & 2779              &   0.077   &   0.465   \\
    \cmidrule{1-1}
    CodeGen2-1B             & 1015              &   0.082   &   0.482  \\
    CodeGen2-3.7B           & 3641              &   0.106   &   0.517   \\ 
    CodeGen2-7B             & 6863              &   0.116   &   0.530   \\ 
    \cmidrule{1-1} 
    CodeParrot-small        & 111               &   0.088   &   0.529   \\ 
    CodeParrot              & 1510              &   0.314   &   0.721   \\ 
    \cmidrule{1-1} 
    InCoder                 & 1312              &   0.115   &   0.559   \\ 
    \cmidrule{1-1} 
    PyCodeGPT               & 111               &   0.079   &   0.567   \\ 
    \cmidrule{1-1}  
    GPT-NEO                 & 2651              &   0.058   &   0.454   \\ 
    \noalign{\smallskip}\bottomrule
    \end{tabular}
    \label{tab:Code_res}
\end{table}

\begin{figure}
    \centering
    \includegraphics[width=0.9\linewidth]{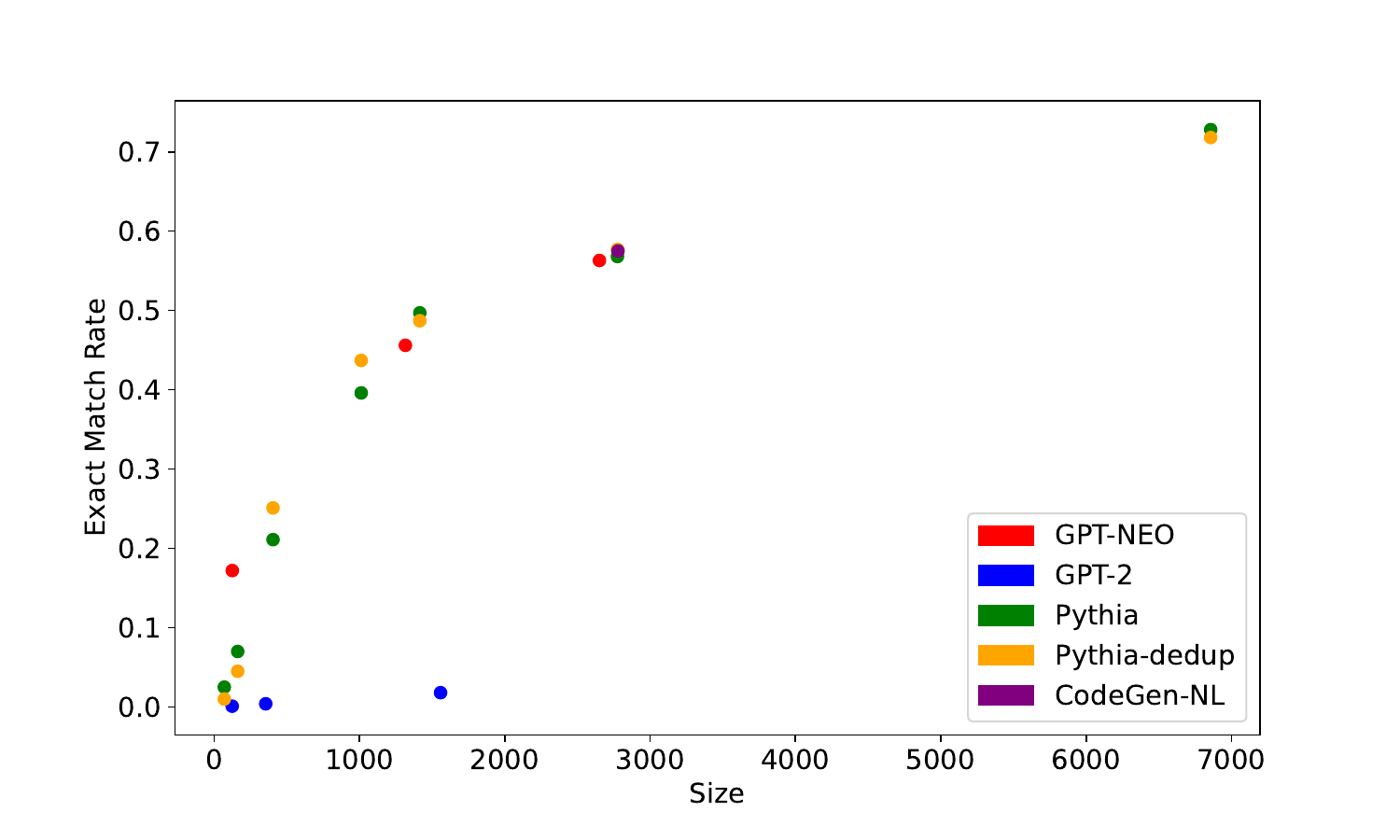}
    \caption{Parameter size and exact match rate for natural language models}
    \label{fig:textEM}
\end{figure}

As shown in~\autoref{fig:textEM}, for the models that are trained on the Pile~\cite{pile}, memorisation scales with the size of the model. We do not observe a clear difference between the Pythia and Pythia-dedup models, indicating that their deduplication was unsuccessful in preventing the memorisation which we measure. As the number of parameters increases for each model architecture, it becomes evident that the rate of memorization grows logarithmically.

\begin{figure}
    \centering
    \includegraphics[width=0.9\linewidth]{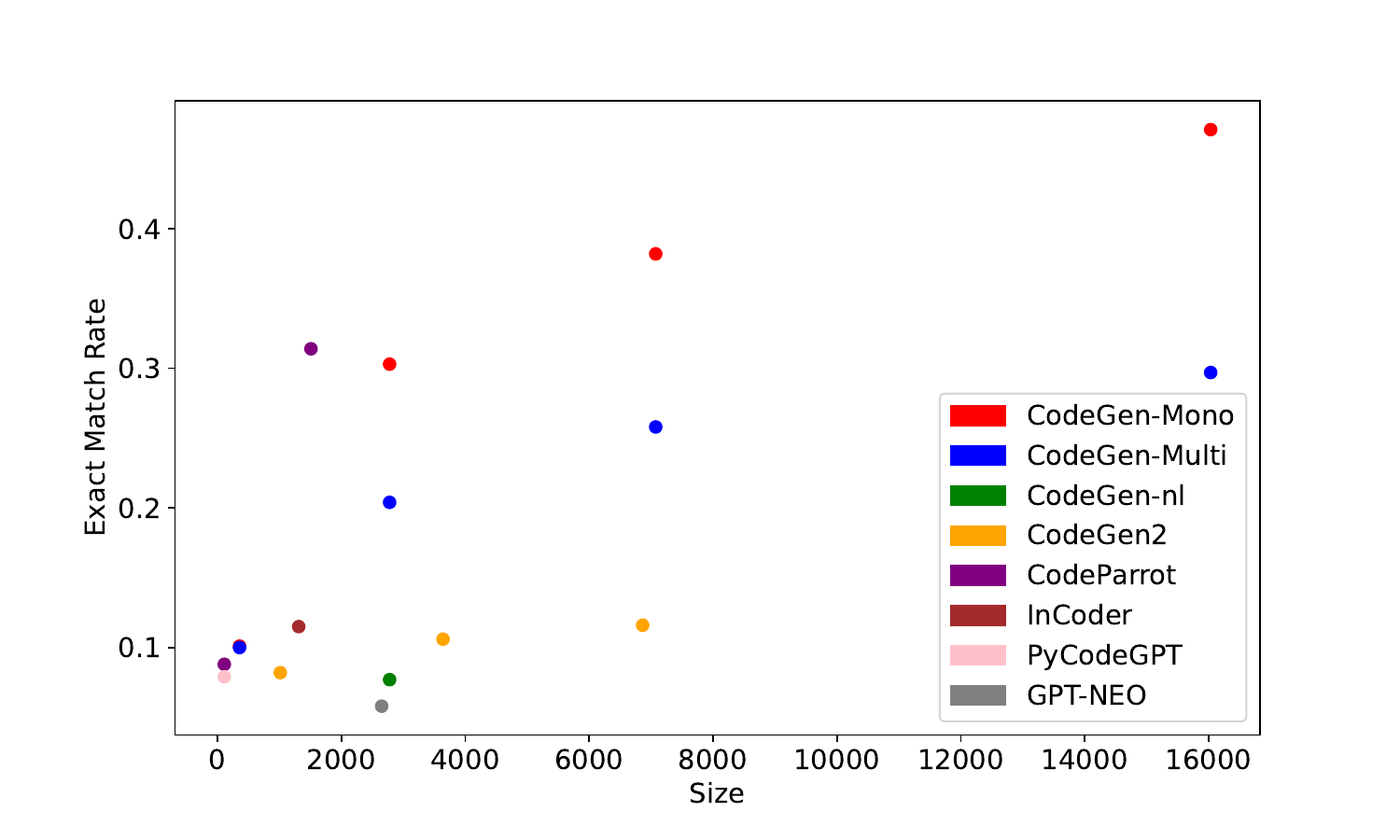}
    \caption{Parameter size and exact match rate for code models}
    \label{fig:codeEM}
\end{figure}

\autoref{tab:Code_res} and \autoref{fig:codeEM} show the results of the experiments. We found that we were able to extract 38\% of the samples from the largest CodeGen-Mono model we tested. The 1B parameter model, which was used to generate the test set, was only able to extract 30\% of the samples, which is lower than the performance of GPT-NEO 1.3B on the natural language dataset. This indicates that our constructed code dataset is harder than the natural language dataset, but that difficulty modifier (2) from \autoref{approach} which was supported by previous works and \autoref{def:mia} also holds for our code dataset.  

\begin{table}
    \centering
    \caption{Natural language attack performance on natural language models}

    \begin{tabular}{lr|cc}
    \noalign{\smallskip}\toprule
    &                   &           \multicolumn{2}{c}{Memorisation rate} \\
    \cmidrule{3-4}
    Model       & Parameters (M)    & EM        & BLEU-4    \\ 
    \cmidrule{1-4}
    GPT-NEO-125M        & 125               & 0.172     & 0.529     \\
    GPT-NEO-1.3B        & 1316              & 0.456     & 0.767     \\ 
    GPT-NEO-2.7B        & 2651              & 0.563     & 0.829     \\ 
    \cmidrule{1-1}
    GPT-2               & 124               & 0.001     & 0.328     \\ 
    GPT-2-Medium        & 355               & 0.004     & 0.375     \\
    GPT-2-Large         & 1558              & 0.018     & 0.396     \\
    \cmidrule{1-1}
    Pythia-70M          & 70               & 0.025     & 0.261     \\
    Pythia-160M         & 162              & 0.070     & 0.355     \\
    Pythia-410M         & 405              & 0.211     & 0.509     \\
    Pythia-1B           & 1012             & 0.396     & 0.658     \\
    Pythia-1.4B         & 1415             & 0.497     & 0.742     \\
    Pythia-2.8B         & 2775             & 0.568     & 0.793     \\
    Pythia-6.9B          & 6857             & 0.728     & 0.880     \\
    \cmidrule{1-1}
    Pythia-dedup-70M    & 70               & 0.010     & 0.273     \\
    Pythia-dedup-160M   & 162              & 0.045     & 0.372     \\
    Pythia-dedup-410M   & 405              & 0.251     & 0.550     \\
    Pythia-dedup-1B     & 1012             & 0.437     & 0.679     \\
    Pythia-dedup-1.4B   & 1415             & 0.487     & 0.712     \\
    Pythia-dedup-2.8B   & 2775             & 0.577     & 0.805     \\
    Pythia-dedup-6.9B   & 6857             & 0.718     & 0.877     \\
    \cmidrule{1-1}
    CodeGen-2B-NL       & 2779             & 0.575     & 0.860     \\
    \noalign{\smallskip}\bottomrule
    \end{tabular}
    \label{tab:NL_res}
\end{table}

\begin{figure}
    \centering
    \includegraphics[width=\linewidth]{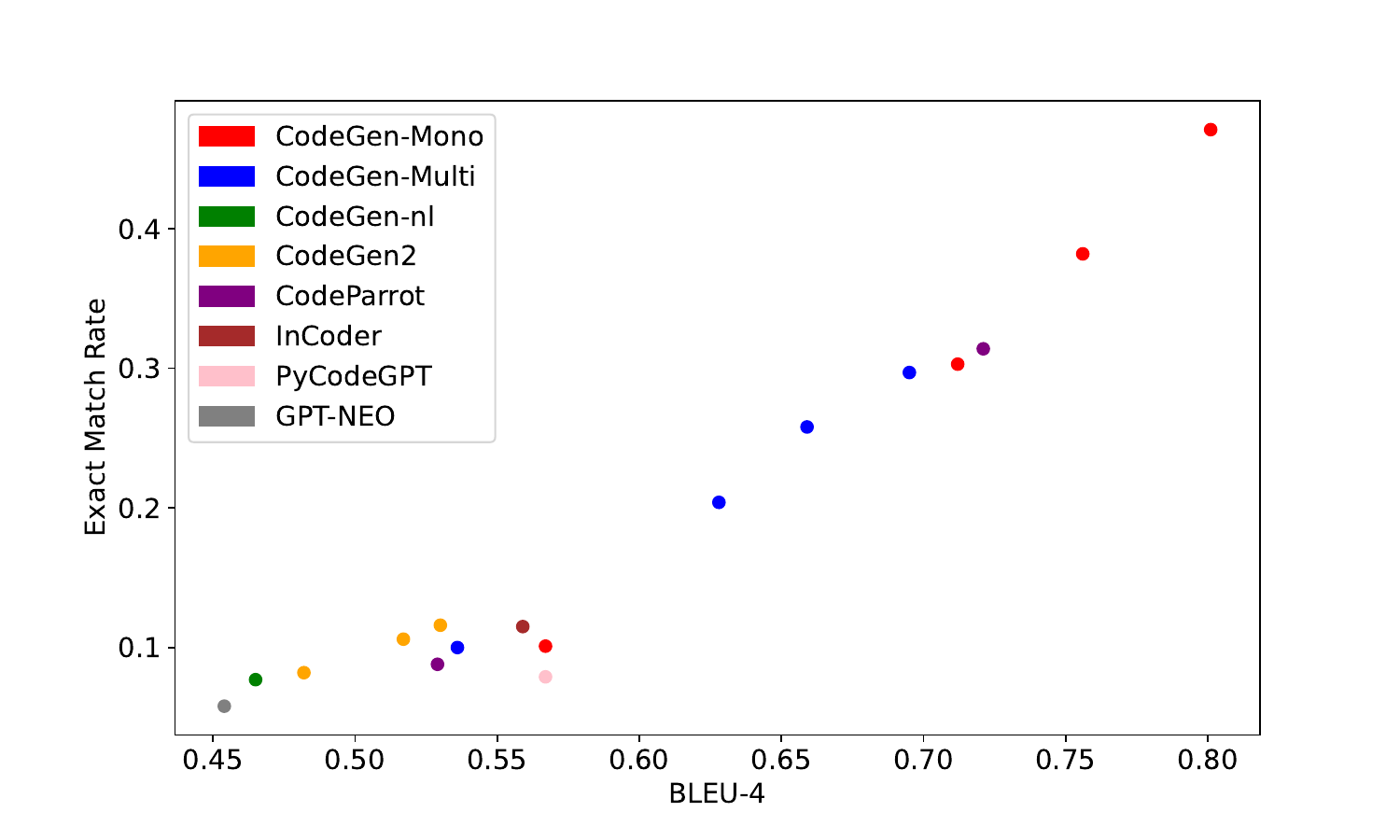}
    \caption{BLEU-4 score and Exact match rate for code models}
    \label{fig:codeEMBleu}
\end{figure}

\autoref{fig:codeEMBleu} shows the relation between the Exact Match rate and the BLEU-4 score for code-trained models. We can observe that there is a clear relation between the exact match rate and the BLEU4 score, especially above an exact match rate of $0.2$. We see a similar pattern in \autoref{fig:codeEMBleu}. The Pearson correlation coefficient between the Exact Match rate and the BLEU4 score is $0.982$ and $0.967$ for natural languageand code, respectively, indicating a very strong positive correlation.

\begin{figure}
    \centering
    \includegraphics[width=\linewidth]{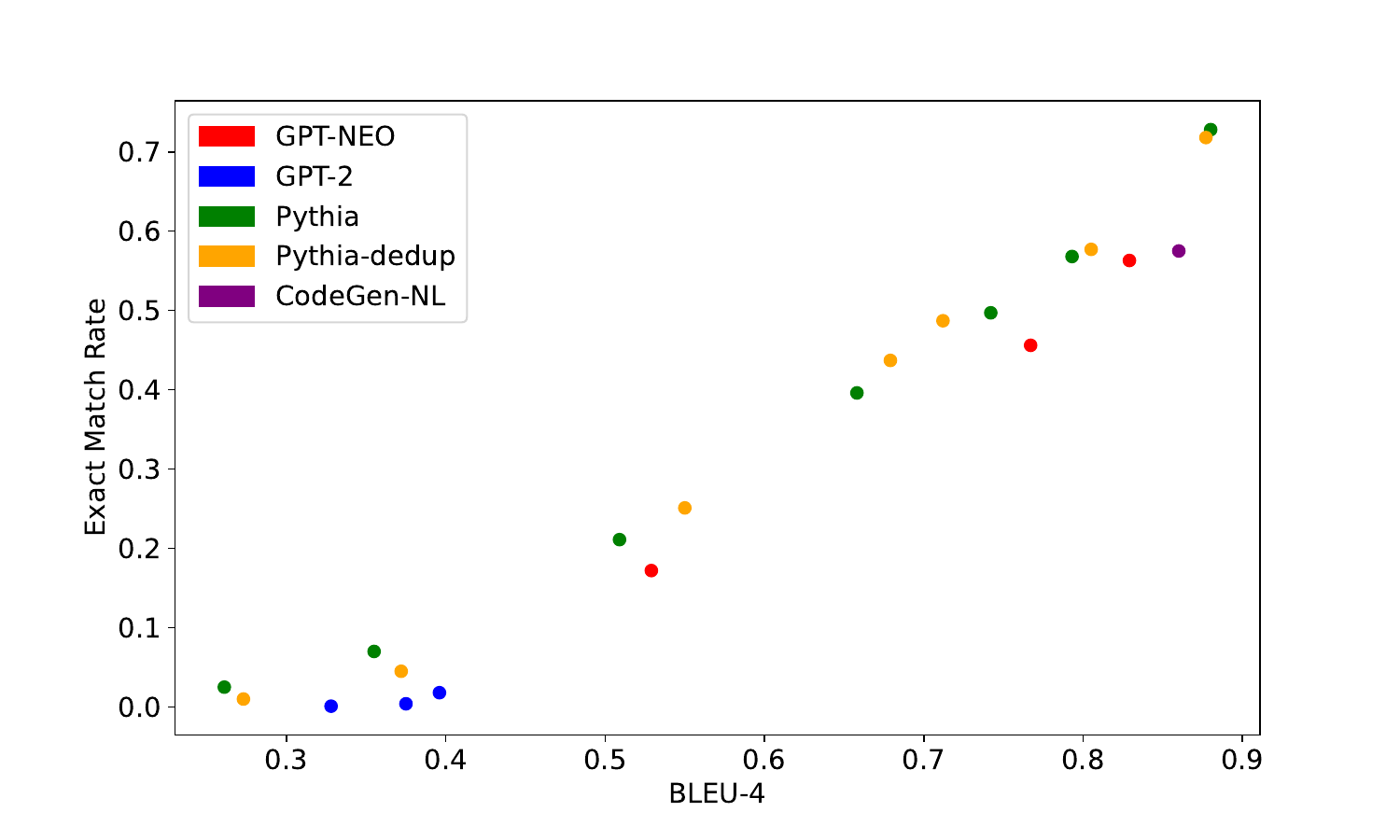}
    \caption{BLEU-4 score and Exact match rate for natural language models}
    \label{fig:textEMBleu}
\end{figure}

In our evaluation, we also tested multiple models that were not primarily trained on programming languages. We found that CodeGen-nl and GPT-NEO were unable to memorise as much as similarly sized code-trained models, but were still able to achieve an exact match score of around 10\%.  

Similarly to natural language models, we also find that memorisation scales with model size in~\autoref{fig:codeEM}. But in this case, we see the logarithmic relationship between the same model architectures. We also observe that the CodeGen-Mono models memorise more natural language than the CodeGen-Multi models for every model size. This indicates that the extra training on Python code increases the memorisation rate. We find a Pearson's correlation coefficient between the Exact Match rate and the size of the model of $0.797$ and $0.704$ for the natural language and the code, respectively, indicating a strong positive correlation.

\begin{RQanswer}
     \textbf{RQ1:} Code-trained LLMs memorise their training data at a lower rate than Natural Language trained LLMs. In both natural language and code-trained models, the rate of memorisation scales with the model size. Continued exposure to the same data increases the rate of memorisation. 
\end{RQanswer}

\subsection{Type of Memorised Samples}
\begin{figure}
    \centering
    \includegraphics[width=0.7\linewidth]{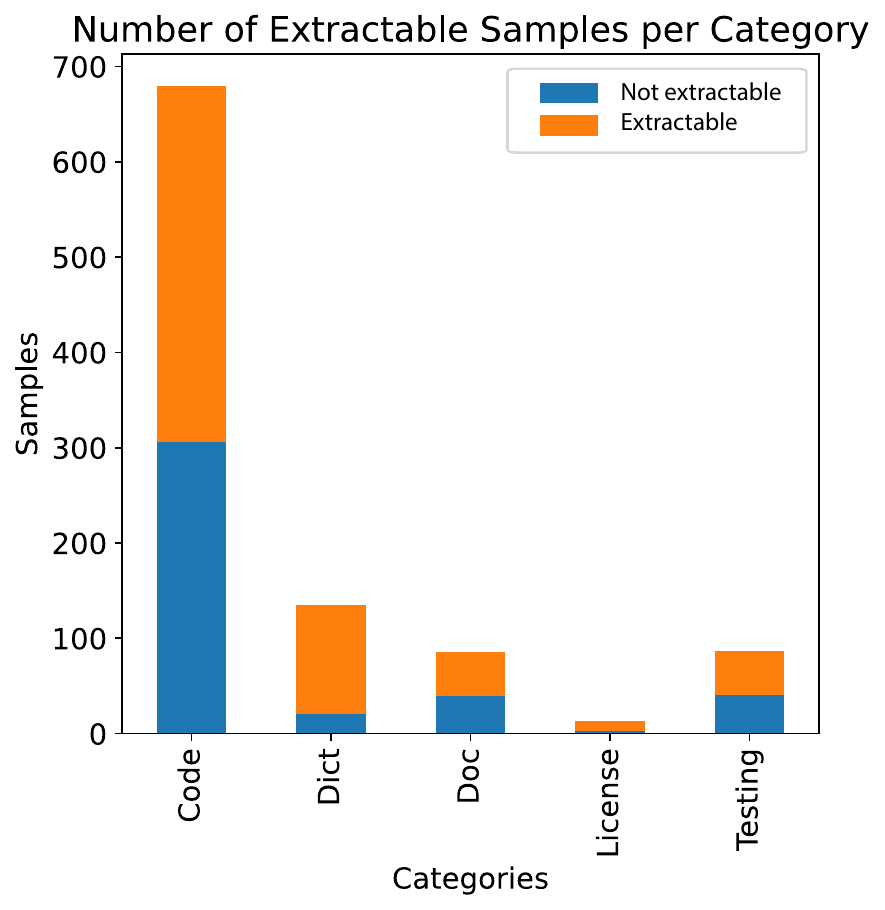}
    \caption{Categories of memorised samples}
    \label{fig:cats}
\end{figure}

As can be observed in~\autoref{fig:cats}, the majority of samples in our dataset are code logic followed by dictionaries. We colour-coded the samples to make a distinction between memorised and non-memorised samples. We find that data carriers and licence information are being memorised at a higher rate than code logic, documentation, and test code.   

During the tagging process, we did find multiple examples of names, emails, and usernames being memorised by the model. Such as the example in \autoref{fig:leakedKey} We also found an example of some API keys, further investigation shows that this instance was a sample that was easily findable using search engines. 

\begin{RQanswer}
     \textbf{RQ2:} LLMs trained on code memorise data carriers and license information at a higher rate than regular source code, documentation, and testing code. Code-trained LLMs are also able to memorise and emit sensitive information. 
\end{RQanswer}

\begin{figure}
    \centering
    \caption{Instance of memorised API keys. Actual keys are replaced with placeholder values.}
    \begin{minted}{python}
{
    'oauth_token: '##############',
    'oauth_token_secret: '################',
    'oauth_verifier: '###########',
}
>>> oauth_session
    \end{minted}
    \label{fig:leakedKey}
\end{figure}

\subsection{Which Model Memorises What}
\begin{figure}
    \centering
    \includegraphics[width=\linewidth]{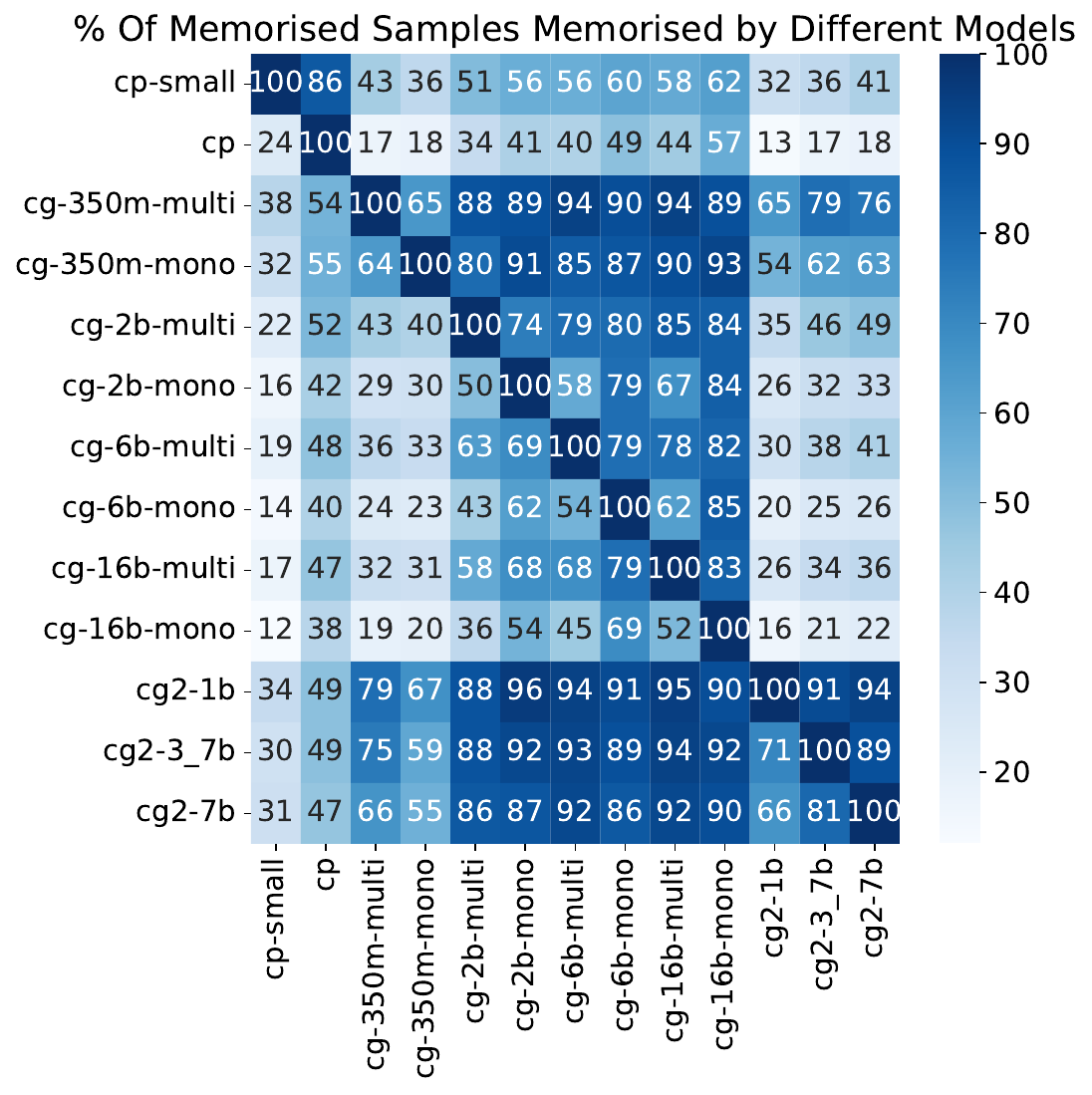}
    \caption{Memorisation overlap between CodeParrot (cp) and CodeGen (cg) Models}
    \label{fig:memOverlap}
\end{figure}

In \autoref{fig:memOverlap} we plot the overlap in memorised samples between different models. We limit the investigation to the Codegen, CodeGen2 and CodeParrot family of models.

For instance, we find that 86\% of all samples which were memorised by CodeParrot-small are also memorised by CodeParrot, while only 24\% of the samples memorised by CodeParrot-small are memorised by CodeParrot. We find similar patterns when comparing the different-sized CodeGen models. The CodeGen-2 family of models memorised fewer samples and is in line with the CodeGen-350M models despite the size difference. The larger models in a family memorise more samples, there are a few distinct samples that are only memorised by the small models, but we find that is generally limited.

\begin{figure}
    \centering
    \includegraphics[width=0.9\linewidth]{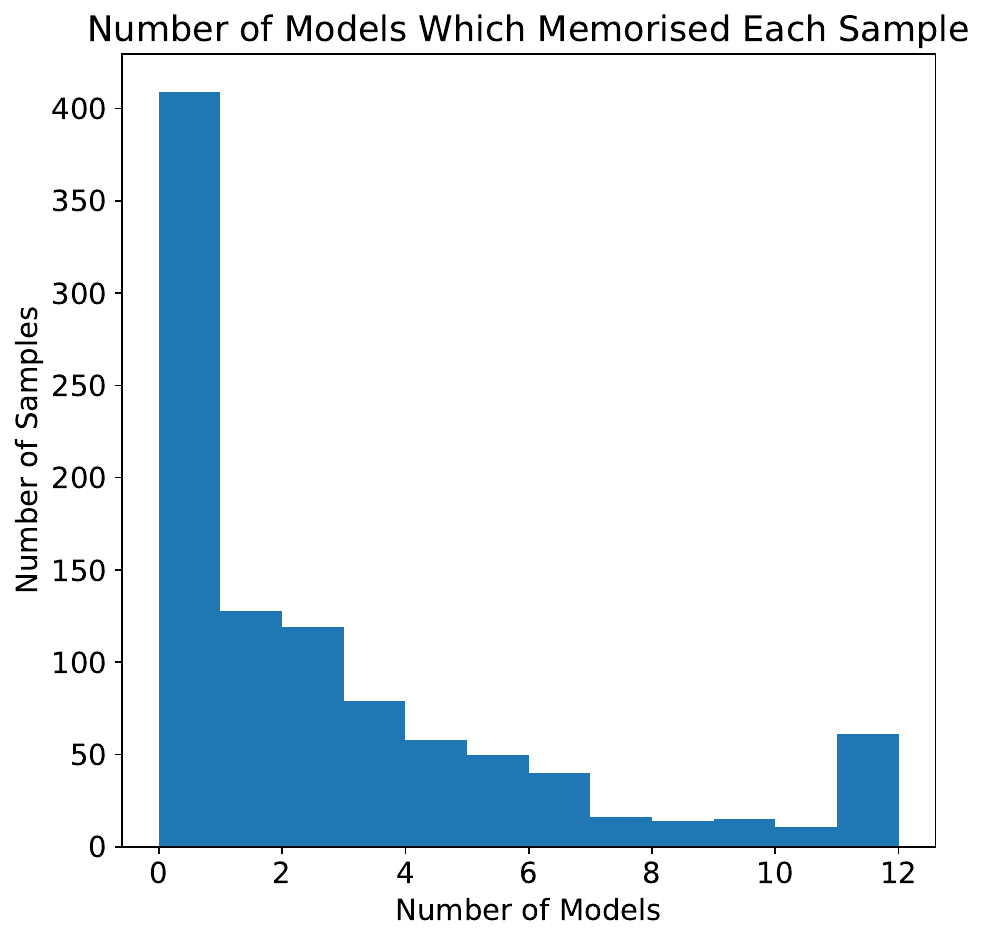}
    \caption{Memorisation counts}
    \label{fig:memCount}
\end{figure}

We find that the CodeGen-Multi models tend to memorise around 50\% of the samples memorised by their respectively sized Mono variant, while the Mono models memorise around 70\% of the samples memorised by the Multi variant. The only exception is the smallest model, where the Multi and Mono models memorised very similar amounts of samples. In ~\autoref{fig:memCount} we find that 40\% of the samples are not memorised by any model at all. But there are 73 samples that are memorised by 12 of all the 13 models. This indicates that there is an inherent difficulty in some samples. 

\begin{figure}
    \centering
    \includegraphics[width=0.9\linewidth]{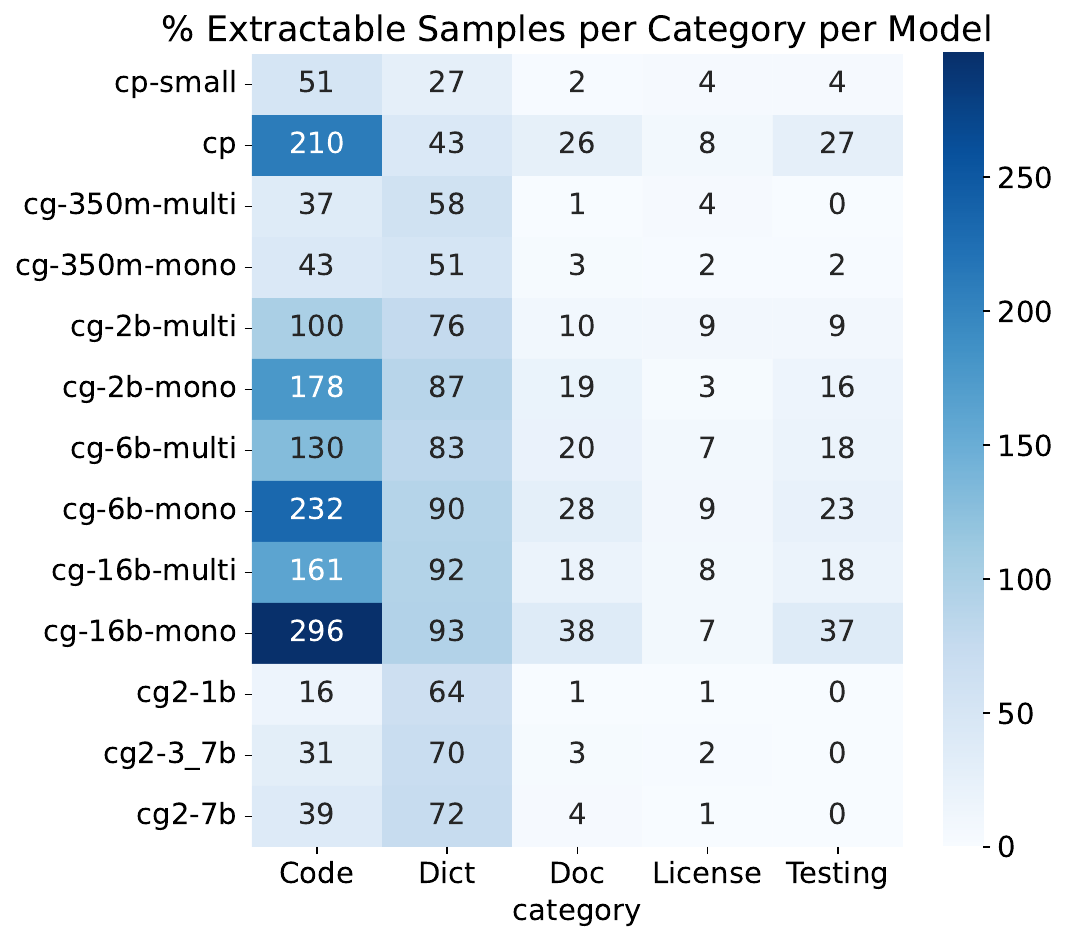}
    \caption{Percentage of extractable samples belonging to each category for CodeParrot (cp) and CodeGen (cg) Models}
    \label{fig:memModels}
\end{figure}

\autoref{fig:memModels} shows the memorisation of each of the categories per model. We find that all plotted models memorise more code and data carriers than any of the other categories, which is supported by ~\autoref{fig:cats}. As models grow larger they memorise relatively more code and fewer data carriers. In absolute terms, the number of memorised samples from the Dict category still increases.

Combined with the findings in RQ1 we can therefore conclude that the extra training on Python, makes the models memorise more and many of the same samples and that the smaller models lack the capacity to memorise more data.

\begin{RQanswer}
     \textbf{RQ3:} Each model family memorises a unique set of samples, and smaller models within the same family remember only a subset of what their larger counterparts do.
\end{RQanswer}

\subsection{Pre-Training Data Leakage}
\begin{figure}
    \centering
    \includegraphics[width=\linewidth]{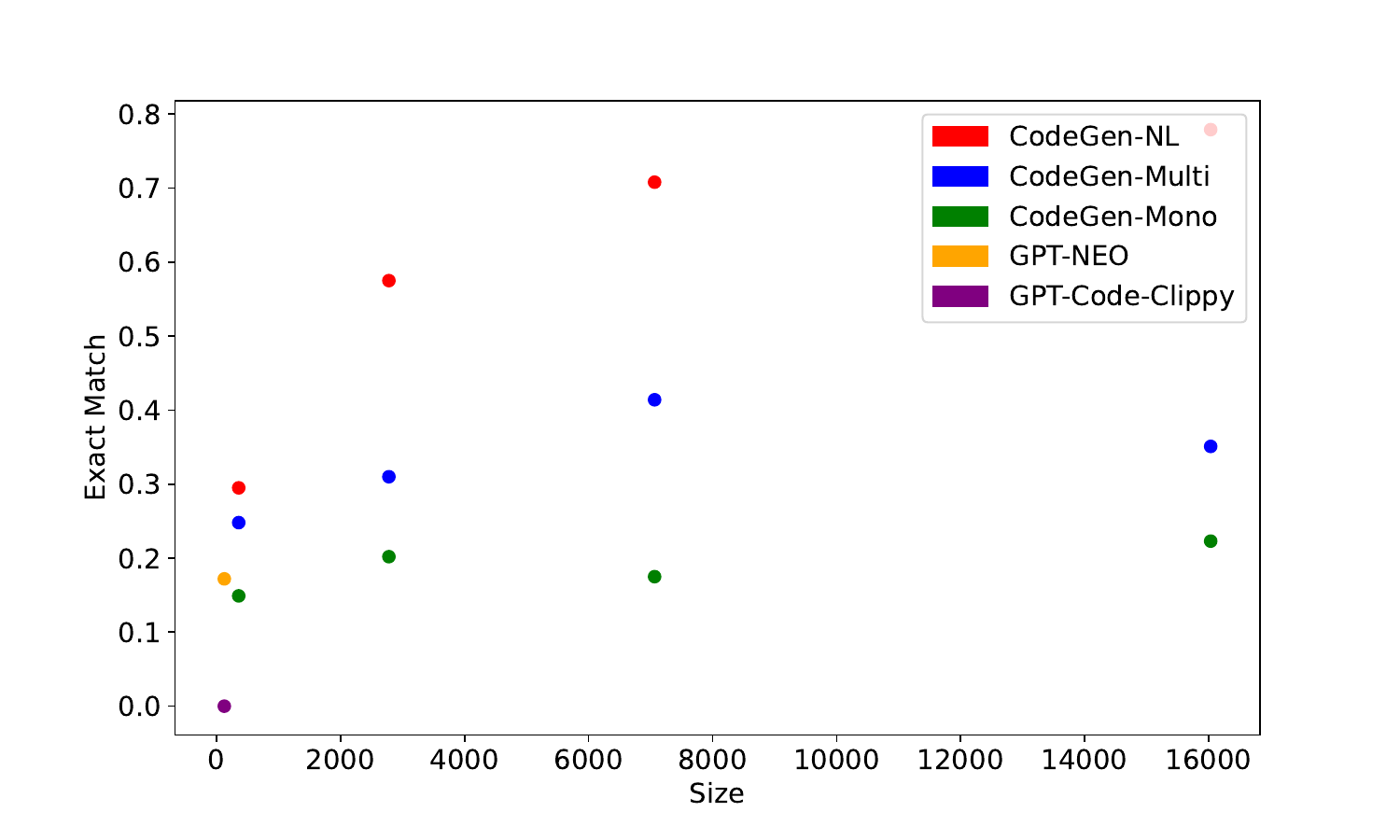}
    \caption{Parameter size and exact match rate for pre-trained models}
    \label{fig:preSizeEM}
\end{figure}

In \autoref{tab:pre_res} and \autoref{fig:preSizeEM} we plot the results for the leakage of pre-training data. We find that we can extract 58\% of all natural language samples from the CodeGen-NL model. This result aligns with the similarly sized Pythia and GPT-NEO models in~\autoref{tab:NL_res}. Tuning the model on code data reduces the extraction rate to 31\% and tuning on Python code further reduces the extraction rate to 20\%. 

We are unable to extract any text samples from GPT-Code-Clippy. The GPT-NEO-125M base model already shows very little extractability in \autoref{tab:Code_res}.

\begin{RQanswer}
     \textbf{RQ4:} While fine-tuning does incrementally reduce the extractability of pre-training data, the pre-training data is still vulnerable to attack, especially as the models grow larger.
\end{RQanswer}

\begin{table}
    \centering
    \caption{Text extraction rate on code models}
    \begin{tabular}{lr|cc}
    \noalign{\smallskip}\toprule
    &                   &           \multicolumn{2}{c}{Memorisation rate} \\
    \cmidrule{3-4}
    Model        & Parameters (M)   & EM        & BLEU-4    \\ 
    \cmidrule{1-4}
        CodeGen-350M-NL & 357 & 0.295 & 0.676 \\ 
        CodeGen-2B-NL & 2779 & 0.575 & 0.860 \\ 
        CodeGen-6B-NL & 7064 & 0.708 & 0.915 \\ 
        CodeGen-16B-NL & 16032 & 0.779 & 0.934 \\ 
        \cmidrule{1-1}
        CodeGen-350M-Multi & 357 & 0.248 & 0.539 \\ 
        CodeGen-2B-Multi & 2779 & 0.310 & 0.588 \\ 
        CodeGen-6B-Multi & 7064 & 0.414 & 0.595 \\ 
        CodeGen-16B-Multi & 16032 & 0.351 & 0.618 \\ 
        \cmidrule{1-1}
        CodeGen-350M-Mono & 357 & 0.149 & 0.454 \\ 
        CodeGen-2B-Mono & 2779 & 0.202 & 0.502 \\ 
        CodeGen-6B-Mono & 7064 & 0.175 & 0.518 \\ 
        CodeGen-16B-Mono & 16032 & 0.223 & 0.546 \\ 
        \cmidrule{1-2}
        GPT-NEO & 125 & 0.172 & 0.529 \\ 
        GPT-Code-Clippy & 125 & 0.000 & 0.148 \\ 
    \noalign{\smallskip}\bottomrule
    \end{tabular}
    \label{tab:pre_res}
\end{table}

\section{Discussion}
\label{discussion}
The results in~\autoref{results} show that large language models pre-trained on source code also memorise their training data and that they are susceptible to targeted training data extraction attacks. 
% Memorisation can have several implications for the measured performance of such models as shown by~\citeauthor{karmakar2022codex}~\cite{karmakar2022codex}. 
\subsection{Interpretation}

\paragraph{Multi vs Mono}
The findings indicate that the CodeGen-Mono models memorised more than the Multi models. This is explainable by the fact that the Mono models have had more exposure to Python code and therefore code in our dataset. Recall that the models are first trained on the Pile which contains all the GitHub repos with more than 100 stars~\cite{pile}. The models are further trained on a general dataset of code, and finally on a dataset of Python code. This means that the models could have possibly been trained on the same file three times. 

\paragraph{Size and Memorisation}
We find that the rate of memorisation scales with the size of the model, across all models we find that the rate of memorisation increases as the size increases. This is in line with the findings of previous work which found that larger LLMs memorise training data faster~\cite{tirumala2022memorization} and at a higher rate than small models~\cite{carlini2022quantifying, carlini2021extracting, carlini2023extractingDiffusion, brown2022does, biderman2023emergent}. Our results also confirm that the log-linear relation between size and memorisation, which has been observed by other works~\cite{carlini2022quantifying, ishihara2023training} holds for LLMs trained on code as well. 

Our experiments which investigate the overlap of memorised sequences in different sizes of code models show that the memorised samples of smaller models are mostly a subset of the large models. This indicates that as a model grows larger it mostly memorises more and not necessarily different data. 

\citeauthor{biderman2023emergent} investigated memorisation in the Pythia suite of models~\cite{biderman2023pythia} and found that 94\% of the sequences memorised by the 70M model were also memorised by the 12B model, but those only accounted for 19\% of the sequences that the 12B model memorised. We find a similar relation between the largest and smallest CodeGen-Mono models: CodeGen-Mono-16B memorised 93\% of the samples which were memorised by CodeGen-Mono-350M, conversely only 20\% of the samples memorised by CodeGen-Mono-16B were memorised by CodeGen-Mono-350M. 

\paragraph{Rate of Memorisation}
Note that the results obtained from experiments in \autoref{results} suggest that memorisation in LLMs trained on code is less than in those trained in natural language. The largest $6.9$B parameter Pythia model memorised 55\% more samples than the best-performing CodeGen-Mono model. Intuitively we would expect the memorisation to be more in code models (as explained in \autoref{methodology}), but there might be multiple reasons for this observation:
\begin{itemize}[leftmargin=*]
    \item Our dataset construction procedure differs from the procedure used by Carlini et. al. The natural language dataset guarantees that for every \(D_n = (s,p)\) there does not exist a \((s,p') \in D\) where \(p \neq p'\). This means that for some prefixes the model might predict a suffix that is also in the training data, which would be counted as a non-memorised sample. This was not possible in our case, since we do not exactly know the training data for the code models under investigation. The training dataset was only deduplicated on the file level.
    \item The structured nature of code might illicit less memorisation in general. This is supported by the higher rate of memorisation in dictionaries compared to regular code especially in smaller models. Their relative information density makes it hard to generalise for these samples specifically and the models might therefore revert to memorisation. 
\end{itemize}

\paragraph{Deduplication}
The deduplicated Pythia~\cite{biderman2023pythia} models are not significantly more robust against our extraction than their regular counterparts. At first glance, this is a surprising finding. It has been reported that deduplicating the training data makes LLMs more secure against data extraction~\cite{kandpal2022deduplicating, carlini2021extracting, lee2022deduplicating}. 

A similar investigation by~\citeauthor{biderman2023emergent} on memorisation on the Pythia suite of models also found a relatively small difference between the two variants~\cite{biderman2023emergent}. The authors theorise that this observation might be due to the training setup. The deduplicated models were trained for 1.5 epochs to offset the smaller data size and to train on the same number of epochs. This effectively oversamples the entire dataset.

Based on our observations we can offer two alternative explanations:
\begin{enumerate}
    \item The training was deduplicated on the file level~\cite{biderman2023pythia}. Our evaluation concerns spans of tokens that can be duplicated across files. The same licence information, for instance, is present in the preamble of many different files and will still be present in the deduplicated dataset.
    \item The samples memorised by the Pythia models might be outliers that illicit memorisation. We observed that information carriers are more likely to be memorised than other types of samples, so the deduplication might not have had much impact on these samples. 
\end{enumerate}

 % While the tested CodeGen models are larger than the GPT-NEO models, our ability to recover training samples is substantially lower. This result needs to be further substantiated by experimenting with more natural language and code models

\subsection{Implications}
\label{future}
% \begin{todolist}
%     \item[\done] Explore lower duplication stats
%     \item[\done] More models (larger, more diverse archs)
%     \item[\done] Optimization of Hyperparams
%     \item[\done] Prompting methods
%     \item[\done] Applying MIA methods from SATML
% \end{todolist}
We propose a novel framework to measure the memorisation and extractability of training data in LLMs. 

\paragraph{Model training}
This work serves to inform researchers and practitioners who aim to train their own LLMs. We can confidently say that larger LLMs leak more and that smaller LLMs are therefore preferable from a safety perspective. In light of emergence~\cite{wei2022emergent}, larger models are however often preferable. We are already able to extract 73\% and 47\% of the text and code samples, even larger models like CodeX~\cite{chen2021evaluating} or Starcoder~\cite{li2023starcoder} might memorise even more data. 

Secondly, we have shown that LLMs also leak their pre-training data even after multiple training rounds. The ability to recover pre-training samples has additional privacy and security implications for the transfer learning paradigm~\cite{alkaswan2023abuse}. When creating and publishing a model, the base model is also something to be considered as the pre-training data can be unintentionally exposed as well. 

Finally, some types of data are more vulnerable to extraction than others. This information can be used to inform the data selection procedure. Some categories like dictionaries can be omitted entirely to reduce the amount of memorisation. Future work can investigate how training data can be curated and sanitised to reduce memorisation in LLMs. 

\paragraph{Model deployment}
The black-box setting of our evaluation has implications for MLaaS services as well. Since we do not require additional information about the model, our data extraction approach could be used against models that are offered through public APIs such as OpenAI's Copilot~\cite{chen2021evaluating}. While Copilot does employ a memorisation filter, it is relatively easy to bypass~\cite{ippolito2022preventing}. There is a need to develop stronger countermeasures to prevent data extraction from these models. 

\paragraph{Framework}
The framework and dataset provided can be used the evaluate different models. While our focus has been on left-to-right causal language models, different architectures, such as encoder-only models like CodeBERT~\cite{feng-etal-2020-codebert} or encoder-decoder models like CodeT5~\cite{wang-etal-2021-codet5} might memorise different amounts and different types of training data.

\paragraph{Fair Use}
Many existing LLMs for code make use of code licenced under copyleft and other non-permissive licences~\cite{alkaswan2023abuse}. The use of public code to train LLMs for code is an instance of fair use, which is a defence that allows the use of copyrighted works in new and unexpected ways and exists in many jurisdictions~\cite{henderson2023foundation}. If the output of the model is similar to the copyrighted input fair use might no longer be applicable. The output needs to conform to the licence terms of the copied input~\cite{henderson2023foundation}, which can include share-alike and attribution clauses~\cite{alkaswan2023abuse}. 

Memorisation can therefore put the creators and users of LLMs for code at legal risk~\cite{henderson2023foundation}. This risk extends to pre-trained models, as some pre-training corpora, including the Pile~\cite{pile}, also contain code licenced under non-permissive licences~\cite{alkaswan2023abuse}. The risk can be avoided by training models with code licenced under permissive licences (such as BSD-3 or MIT) or providing provenance information to trace the code back to its source so that the user of the output can abide by the original licence~\cite{henderson2023foundation, li2023starcoder}. 

\paragraph{Extraction techniques}
We were able to show that using relatively simple greedy decoding and the notion of k-extractability, most text models and all code models are leaking data. This only proves the inherent leakiness of these models and serves as a stepping stone for more advanced and powerful attacks. 
One approach worth investigating is the use of prompt engineering to extract data. With hard or soft-prompts~\cite{liu2023pre} the model could be enticed to output more memorised data. Our work only prompts the models with the prefix, while different prompts might elicit more memorisation. 
Another approach is to explore the use of Membership Inference Attacks to increase the abilities of the attacks further. One could take inspiration from untargeted attacks and generate multiple suffixes per prefix using a different decoding method. The MIA can then serve to select the correct suffix~\cite{alkaswan2023extending}.

\subsection{Limitations and Threats to Validity}
\label{threats}

\subsubsection{Internal validity} In our evaluation, we did not take into account the location of the samples. The samples are of a fixed token length but can originate from any arbitrary location in the file. Furthermore, Byte-Pair Tokenisation can cause the sample to start or end in the middle of a word. We based our dataset construction on existing work~\cite{biderman2023emergent, alkaswan2023Targeted}, but samples from the beginning or end of the file could be easier to extract. Initially, untargeted extractions were attempted, and it was discovered that samples were predominantly obtained from the beginning of the file. Nevertheless, the current approach was chosen as it would enhance the versatility of our attack and enable us to extract samples from any location within the file.  

\subsubsection{External validity} Our evaluation focuses on a limited number of models, other models might exhibit more or less memorisation. Our benchmark was constructed using a single model, and while we were able to show that our benchmark gave promising results for other models, other data sources and models should be used to construct more benchmarks. 

The constructed datasets only consider duplicated sequences; this inherently limits the applicability of our attack on low-duplication data. While other works do state that models can also memorise unduplicated data, we cannot experimentally confirm this as we only apply coarse file-level deduplication. 

In the construction of our dataset, we only considered Python code. We selected Python because it is supported by almost all code generation models. Other less-expressive languages could show different patterns and different degrees of extractability. Python is a very popular language, so these results might also not apply to less popular languages. We plan to extend our evaluation to include more programming languages in the future.

\subsubsection{Construct validity} We mainly use the exact match metric to measure memorisation in code models. This metric likely underestimates the actual number of memorised samples, as some might be slightly changed by the model. For this specific study, we are more interested in exact reproductions by the model, since we are more interested in the privacy and security aspect of memorisation. When examining the licensing aspects of memorization, fuzzy match metrics may provide better insights. We included BLEU4 to account for this, but we found that it is highly correlated with the exact match rate. However, there are no automated metrics available to measure non-literal infringement based on current legal standards~\cite{henderson2023foundation}.

\subsubsection{Ethical Considerations}
While this work does describe techniques that can potentially be used to extract sensitive information from models, we do so ethically. Our goal is to bring attention to the issue of memorisation in LLMs for code and inform the users and creators of these models and provide them with tools to measure this. In this work, we, therefore, do not needlessly expose any private information, and we urge users of our framework to refrain from doing so as well. We target randomly selected sequences from popular and public repositories to avoid accidentally exposing private information. However, we still found some instances of usernames, emails, and API keys in our data, but we found that these are easily findable using search engines and are part of popular and well-indexed public repositories. We believe that the benefits outweigh the risks, and we decide to share our datasets. 

\section{Conclusion}
% what we did
To conclude, we presented an extensive study on memorisation in LLMs for code. We formally define a data extraction security game grounded in the existing notion of k-extractability and membership inference attacks. We utilised this game to create a dataset to measure memorisation in LLMs for code. We compared the rate of memorisation between models of code and natural language, we compared the rate and type of memorisation between different models, and we investigated the rate of memorisation of pre-training data in LLMs for code.

%Findings
We found that LLMs for code memorise their training data like their natural language counterparts, albeit at a lower rate. We further found that the rate of memorisation increases as a model grows and that different model architectures memorise distinct sets of samples, while smaller versions of the same family tend to memorise a smaller subset of their larger sibling. We found that data carriers and licence information are being memorised at a higher rate than code, documentation, and tests. Finally, we found that the pre-training data is still vulnerable to extraction even after multiple tuning rounds. 

% Recommendation
Our work is a first step and provides a framework to measure memorisation in LLMs for code. We strongly advise the research community to conduct a more comprehensive investigation into the extent of data leakage and employ a diverse range of models and extraction techniques to develop safeguards that can effectively mitigate this issue. The consequences of data leakage can be severe, so it is crucial to take proactive measures to address this problem.

\newpage
\bibliographystyle{ACM-Reference-Format}
\bibliography{references.bib}

\end{document}